\documentclass[amsmath,twocolumn,superscriptaddress]{revtex4-1}
\usepackage{epsfig}
\usepackage{amssymb}
\usepackage{amsmath}
\usepackage{amsfonts}
\usepackage{tcolorbox}
\usepackage[T1]{fontenc}
\usepackage{bm}
\usepackage{graphicx}
\usepackage{color}
\usepackage{verbatim}

\begin{document}


\title{Quantum Order-by-Disorder in Strongly Correlated Metals}

\author{Andrew~G. Green}
\affiliation{London Centre for Nanotechnology, University College London, Gordon St., London, WC1H 0AH, United Kingdom}

\author{Gareth Conduit}
\affiliation{TCM, Cavendish Laboratory, JJ Thomson Ave, Cambridge, CB3 0HE}

\author{Frank Kr\"uger}
\affiliation{London Centre for Nanotechnology, University College London, Gordon St., London, WC1H 0AH, United Kingdom}
\affiliation{ISIS Facility, Rutherford Appleton Laboratory, Chilton, Didcot, Oxfordshire, OX11 0QX, United Kingdom}

\begin{abstract}
Entropic forces in classical many-body systems, e.g. colloidal suspensions, can lead to the formation of new phases. Quantum fluctuations can have similar effects: spin fluctuations drive the superfluidity of Helium-3 and a similar mechanism  operating in metals can give rise to superconductivity. It is conventional to discuss the latter in terms of the forces induced by the quantum fluctuations. However, focusing directly upon the free energy provides a useful alternative perspective in the classical case and can also be applied to study quantum fluctuations. Villain first developed this approach for insulating magnets and coined the term order-by-disorder to describe the observed effect. We discuss the application of this idea to metallic systems, recent progress made in doing so, and the broader prospects for the future. 
\end{abstract}

\maketitle

\tableofcontents

\section{INTRODUCTION}
The entropic generation of forces is familiar in classical systems. Examples include the elastic forces in stretched rubber and the apocryphal attraction between ships in a swell. Such forces arise due 
a state-dependent restriction of the
spectrum of fluctuations and thus their entropic contribution to the free energy.  In many body systems, such forces can lead to phases that are favoured for entirely entropic reasons. In colloids, they are responsible for a variety of different phases with characteristic dependence upon the shape of the colloidal particles\cite{onsager1949effects,Adams:1998kq}.
Similar effects are responsible for the folding of DNA due to the conformational entropy of the surrounding water, and the celebrated Berezinskii-Kosterlitz-Thouless transition\cite{BerezinskiiI,BerezinskiiII,KTtransition} is driven entirely by the entropy of unbound vortices. 

Often, it is revealing to discuss these transitions directly at the level of the free energy rather than through the resulting forces. Viewed in this way, the idea of entropically driven order is a powerful unifying concept. In engineering it comes under the banner of the term state-dependent noise as exemplified by the noisy inverted pendulum\cite{MeersonInvertedPendulum}. Further afield, in business or evolution, fluctuations may make adaptability a favourable strategy. Essentially, fluctuations can stabilise financial or ecological niches that would not be stable in their absence. Entropic priors in Bayesian inference bias data analyses in 
the same way\cite{mackay2003information}. 

{\it Quantum fluctuations} can generate forces in a very similar manner to classical entropic effects, the Van der Waals force being the most famous example\cite{LondonVdW,casimir1948attraction}. Forces generated by the fluctuations of quantum spins are responsible for the superfluidity of Helium-3\cite{anderson1973anisotropic,brinkman1974spin} and a similar mechanism operating in metals can give rise to spin-fluctuation induced superconductivity\cite{fay1980coexistence}.
In all of these cases, it is conventional to discuss the effect of quantum fluctuations in terms of the forces that they induce. However, focussing directly upon the free energy  provides a useful  alternative perspective. In the quantum case, it is the zero-point energy of fluctuations rather than an entropic contribution to the free energy that is at play, however, the effects are very similar. 
Indeed, the terms Casimir force and Van der Waals force are sometimes generalised beyond their original context of forces due to zero-point fluctuations of the electromagnetic vacuum, to refer to entropic forces in colloids. 

Villain provided a concrete example of the utility of focussing directly upon the free-energy\cite{villain1980order} and coined the term {\it order-by-disorder} to describe the observed effect. He considered a magnetic model whose classical groundstates form a degenerate manifold. Allowing for fluctuations (either quantum zero-point or thermal\footnote{
The very similar effects of zero-point fluctuations and thermal fluctuations is revealed particularly clearly; a factor of $n_{\hbox{B}}+1/2$  for each spinwave mode accounts for both zero-point (as $T\rightarrow 0$) and thermal (as $T\rightarrow \infty$) occupation. 
}) 
about these classical configurations breaks the degeneracy and picks out a particular ordered state --- hence the term order-by-disorder. 
These ideas  have been applied in a number of insulating magnets, especially in cases where frustration leads to a degenerate manifold of classical groundstate configurations that is broken by  fluctuations, though this degeneracy is not necessary for the fluctuation contribution to the free energy to be appreciable. Examples include spin ice pyrochlores\cite{MoessnerOBD1,MoessnerOBD2} and frustrated antiferromagnets on the honeycomb lattice\cite{ParamekantiOBD}.

The main focus in this perspective is the application of order-by-disorder to metallic systems. The central philosophy  is simple to state: modifications of the Fermi surface, for example by the introduction of some order parameter, modify the electron dispersion and as a result reconfigure the spectrum of low-lying 
excitations. This in turn shifts the zero-point energy (and entropy at finite temperature) of fluctuations. When fluctuations are large, this may self-consistently determine the Fermi surface. 
The similarity of this to the Casmir effect is clear, as is the importance of fermionic statistics. The Fermi surface self-consistently provides boundary conditions for the electrons {\it via} Pauli exclusion. 
One could say that particle-hole excitations induce generalised Casimir or Van der Waals forces between electrons. 
We first present these ideas in the context of the critical itinerant ferromagnets, where it was initially developed.  We give both a heuristic presentation and an overview of a more formal field theoretical derivation, emphasising the close parallel with Villain's order-by-disorder. 

As in the classical case, the order-by-disorder approach reveals commonalities between effects that are difficult to appreciate from other perspectives. This can be particularly advantageous when trying to understand experiments. Measurement of the fluctuation spectra and appreciation of how it is altered by different types of order can lead directly to predictions of what types of instability a system is prone to, even in the absence of a detailed microscopic model. We discuss several developments of fermionic quantum order-by-disorder that  were directly influenced by experiment.

The difference between spin-fluctuation theory focussing on forces, and  fermionic quantum order-by-disorder focussing upon the free energy is one of perspective. Both encompass the same physics and, in circumstances where direct comparisons can be made, lead ultimately to the same equations. 
However, there are other methods --- especially numerical --- that can be used to analyse strongly correlated quantum systems. 
We illustrate the relationship that fermionic quantum order-by-disorder bears to them;
its comparison to various {\it ab initio} techniques, the potential for future inclusion in density functional code, and agreement with Monte Carlo calculations for the itinerant ferromganet. Finally, we discuss the prospects for near- and long-term development of the technique and how the broader application of the idea has resonance with ideas of fluctuation-induced geometry of entanglement structure and entropic gravity. 

\section{THE FERROMAGNETIC METAL }
\label{sec:Ferromag}
\subsection{Simple Model and Historical Development}
\begin{widetext}

\begin{figure}
\includegraphics[width=0.6\textwidth]{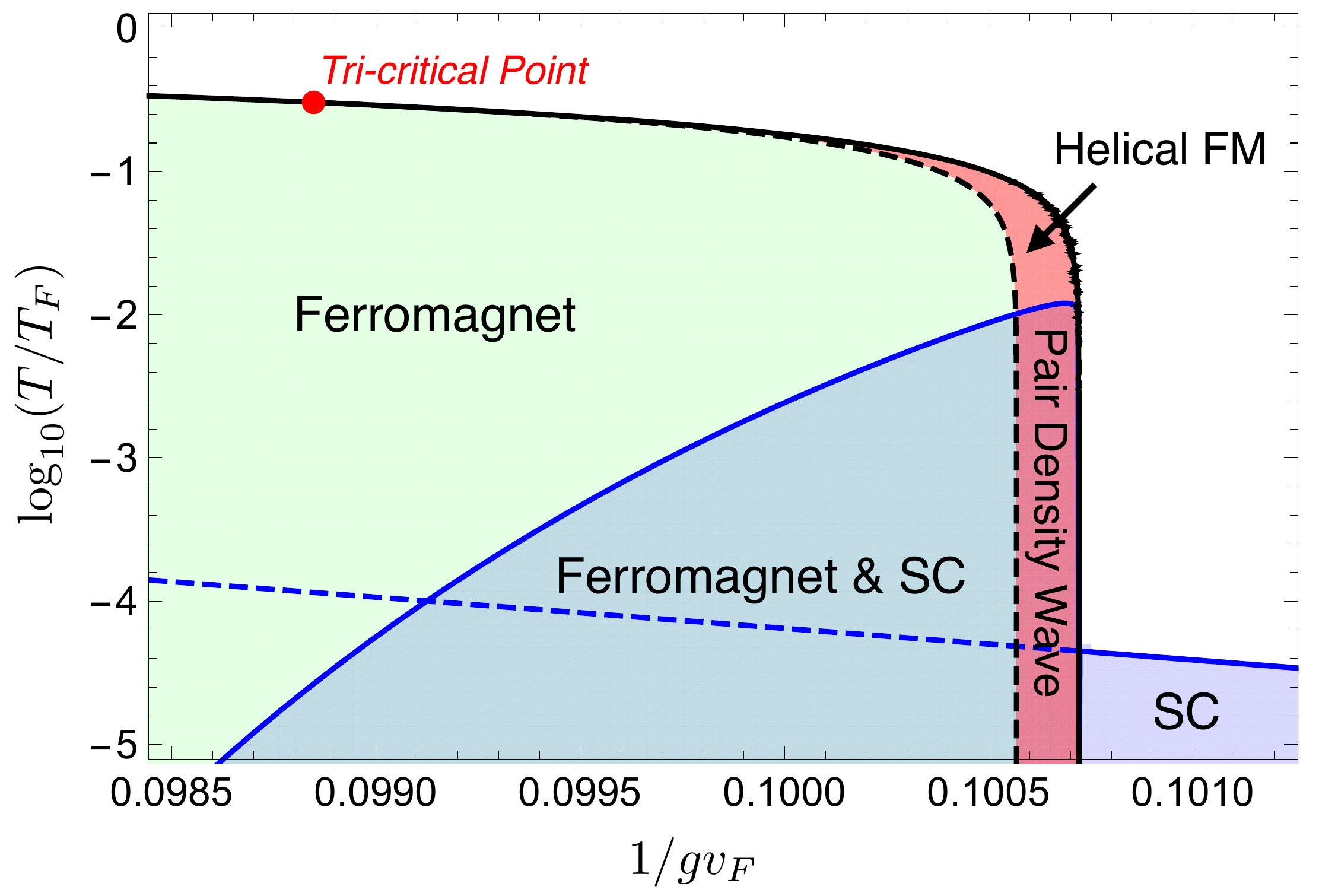}
\caption{
{\it Phase Diagram of the Critical Ferromagnet:} Quantum fluctuations in the vicinity of the putative quantum critical point drive a reconstruction of the phase diagram. In the vicinity of the tricritical point, where the transition becomes first order, an expansion of the Ginzburg-Landau function to quartic order is sufficient. Approaching zero temperature, the phase boundary is determined by singularities in the higher order terms of the expansion. Fluctuations drive a superconducting pairing in the $p$-wave channel by the same mechanism as the pairing in $^3$He. A novel pair density wave order is found in the region of the phase diagram where superconductivity and helimagnetic order overlap.}
\label{fig:PhaseDiagram}
\end{figure}

\end{widetext}
The simplest non-trivial fermionic model to which one can apply  quantum order-by-disorder is that of electrons with a quadratic dispersion, $\epsilon_{\bf k} = {\bf k}^2$, and contact interaction, $g$;
\begin{equation}
{\cal H}
=
\sum_{{\bf k},\sigma} 
\epsilon_{\bf k} \hat c^\dagger_{{\bf k} ,\sigma} \hat c_{{\bf k} ,\sigma} 
+ 
g \int d^3 {\bf x}
\;
 \hat c^\dagger_{{\bf x} ,\uparrow} \hat c^\dagger_{{\bf x} ,\downarrow}
 \hat c_{{\bf x} ,\downarrow} \hat c_{{\bf x} ,\uparrow},
\label{TheHamiltonian}
\end{equation}

with  $\hat c^\dagger_{{\bf x}({\bf k}), \sigma}$ a fermionic creation operator in position (momentum) space. Below, we take the spin label $\sigma=\pm$ corresponding to up- and down-spin relative to the local magnetisation. 
Despite its apparent simplicity, this model displays a remarkable range of different phenomena. Moreover, it finds direct realisation in cold atomic gases where atomic interactions are local. It is a good approximation to electrons in solids with a chemical potential near the bottom of a band so that the dispersion is approximately quadratic and where screening renders the effective Coulomb interaction between electrons short ranged --- we shall discuss the effects of longer range interactions later. 

A mean-field analysis allowing for the possibility of a finite 
magnetization, $M=\sum_{\sigma=\pm} \sigma \langle \hat n_{{\bf x}, \sigma} \rangle$,  gives rise to the free energy
\begin{equation}
{\cal F}_{\text{MF}}
=
- \frac{1}{\beta} \sum_{{\bf k},\sigma} \ln \left( 1+ e^{-\beta(\epsilon_{\bf k}-\sigma g M - \mu)} \right)
+ g \int d^3{\bf x} M^2({\bf x}).
\label{MeanFieldFreeEnergy}
\end{equation}
This is simply the Stoner model of ferromagnetism. It shows a second order transition between paramagnetic and ferromagnetic phases, the temperature of which varies with the interaction strength $g$ and which occurs at $\rho_{\hbox{F}} g=1$ at zero temperature ($\rho_{\hbox{F}}$ being the density of states at the Fermi level). 

In his seminal paper on this model, Hertz\cite{hertz1976quantum} realised that fluctuations in certain regions of the phase diagram are profoundly affected by  quantum mechanics. This leads to behaviour in a different universality class to that of classical phase transitions; a class that Hertz termed quantum critical. Hertz extended the free energy given in Eq.(\ref{MeanFieldFreeEnergy}) to an action for the dynamical magnetization field, the dynamics of which is given by the decay of magnons into particle-hole pairs, {\it i.e.} Landau damping\cite{landau1946vibrations}. This modification --- and the assumption that the non-analyticity of the Landau damping does not extend to other terms under renormalization --- yields  power-law dependence of physical quantities on temperature that are characteristic of the quantum critical point\cite{hertz1976quantum,moriya2012spin,millis1993effect}. 
The presence of strong ferromagnetic fluctuations near to the zero-temperature limit of the phase transition can also lead to superconducting pairing in the p-wave channel, {\it via} a mechanism translated from superfluid\cite{anderson1973anisotropic,brinkman1974spin}  $^3$He to metallic systems by Fay and Appel\cite{fay1980coexistence}.

\begin{figure}
\includegraphics[width=0.2\textwidth]{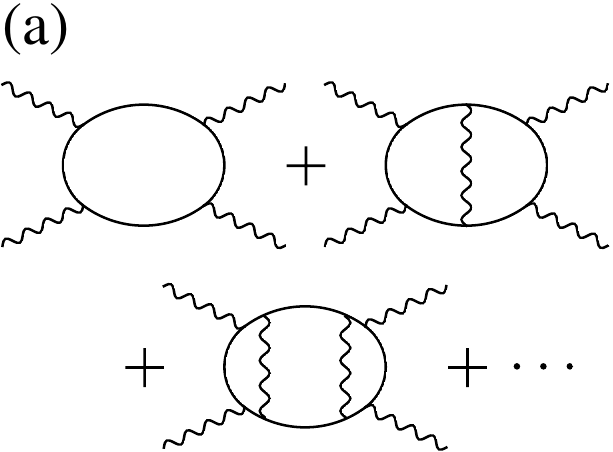}\\
\includegraphics[width=0.3\textwidth]{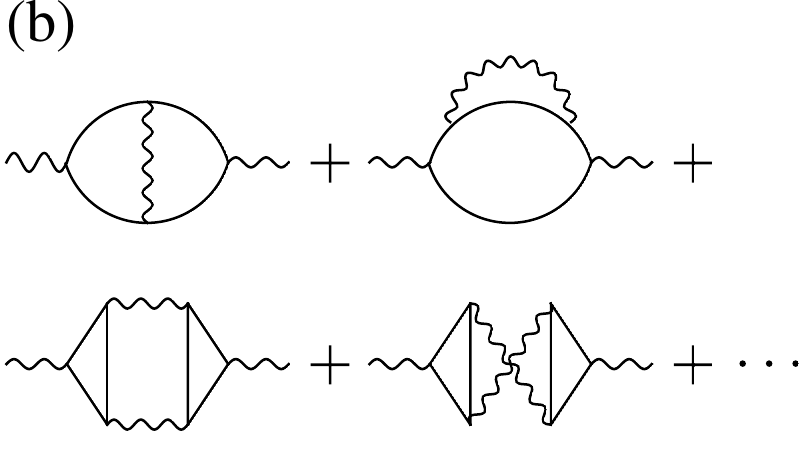}
\caption{
{\it Diagrams contributing to non-analytic extensions to Moriya-Hertz-Millis theory:} Re-summation of diagrams of these types propagate non-analyticity from Landau damping to all diagrams in the expansion. These ultimately lead to reconstruction of the phase diagram in the vicinity of the quantum critical point.  (a) Contributions to non-analyticity in the order parameter.
(b) Contributions to non-analyticity in gradients of the order parameter. }
\label{fig:NonanalyticDiagrams}
\end{figure}
However, this story is not complete.  It was first realised by Belitz, Kirkpatrick and Voijta\cite{belitz1997nonanalytic} that the Moriya-Hertz-Millis theory suffers from an internal inconsistency\footnote{In fact, the existence of these non-analyticities had been noted earlier\cite{PhysRevB.15.1523} but their physical significance was not understood until the work of Belitz-Kirkpatrick-Vojta.} --- the non-analyticity present in Landau damping propagates under renormalization to all terms in an expansion of the action\cite{belitz1997nonanalytic,betouras2005thermodynamics,rech2006quantum,efremov2008nonanalytic,maslov2009nonanalytic}.  
This presages the fact that fluctuations in the vicinity of the quantum critical point are in fact so strong that they favour a reconstruction of the phase diagram in the vicinity of the quantum critical point.
The Ginzburg-Landau function for the magnetization incurs non-analyticities in both the magnetization (of the form $M^4 \log M$ in three dimensions and $M^3 \log M$ in two dimensions) and its gradients (of the form $-|{\bf q}| M^2$ in three dimensions and $-|{\bf q}|^{3/2} M^2$ in two dimensions). These drive the transition into the ferromagnet first order and favour helimagnetic order at low temperatures in the vicinity of the quantum critical point. 

It is natural to ask whether the first order ferromagnet, helimagnet, and superconducting instabilities have a common cause. It is not at all obvious from the diagrammatics that they do [See Fig.\ref{fig:NonanalyticDiagrams}]. Moreover, it would be desirable to have a simple heuristic picture that allows one to anticipate the helimagnetic instability in advance of detailed calculation. As we shall see next,  fermionic quantum order-by-disorder provides this unified description and has a simple heuristic interpretation. 

\subsection{Fermionic Quantum Order by Disorder}

\subsubsection{Heuristic Description}
\begin{widetext}

\begin{figure}
\includegraphics[width=0.85\textwidth]{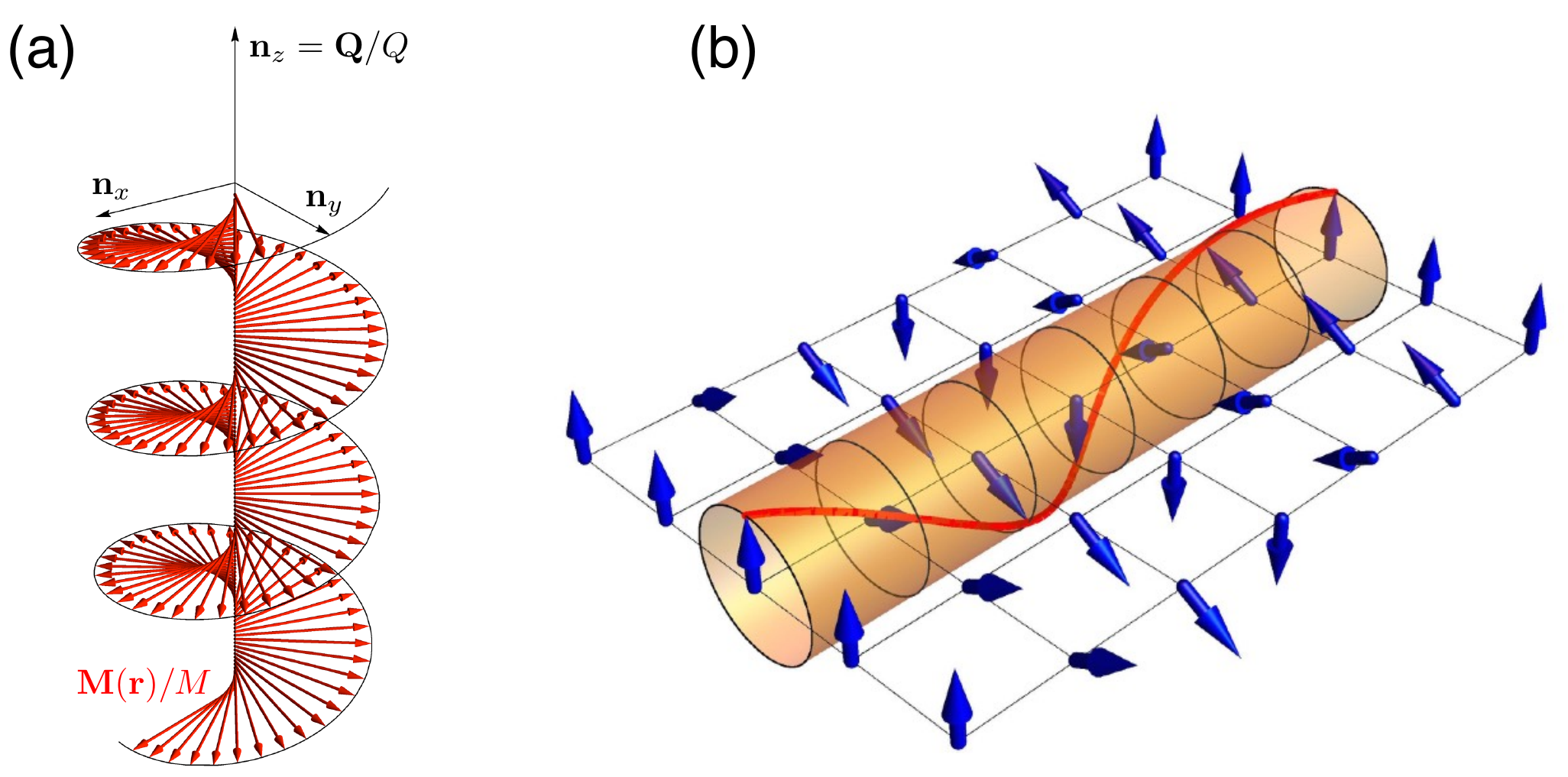}
\caption{
{\it Illustration of Helimagnetic Order:} In the vicinity of the itinerant ferromagnetic quantum critical point, fluctuations drive the formation of helimagnetic order in which --- as illustrated in (a) --- the magnetic quantization axis rotates when moving along the direction of the pitch vector ${\bf q}$. A realisation of this in the square lattice is shown in (b).}
\label{fig:Helimagnet}
\end{figure}
\end{widetext}

The essence of fermionic quantum order-by-disorder is to combine mean-field and fluctuation (zero-point or entropic/thermal) contributions to the free energy whilst self-consistently allowing for the possibility of additional order, such as helimagnetic or superconducting order. When the fluctuation contributions are large, they may provide the dominant contribution to the free energy and determine the state adopted by the system. In the case of the simple model given in Eq.(\ref{TheHamiltonian}), the leading fluctuation correction to the free energy is given by
\begin{equation}
{\cal F}_{fl}
=
-2 g_{\hbox{eff}}^2 
\sum_{{\bf k}_1...{\bf k}_4}' 
\frac{ 
f^+_{{\bf k}_1}
f^-_{{\bf k}_2}
(f^+_{{\bf k}_3} +f^-_{{\bf k}_4})
}{\epsilon^+_{{\bf k}_1}+\epsilon^-_{{\bf k}_2}
-\epsilon^+_{{\bf k}_3}-\epsilon^-_{{\bf k}_4}},
\label{FluctuationCorrection}
\end{equation}
where the summation is taken over momenta such that ${\bf k}_1 + {\bf k}_2={\bf k}_3+{\bf k}_4$. The electron dispersion for spin $\sigma$ is given, for example in the presence of helimagnetic order [See Fig.\ref{fig:Helimagnet}] with pitch vector ${\bf q}$, by $\epsilon^\sigma_{{\bf k}}= \epsilon_{\bf k} - \sigma \sqrt{ ({\bf q}\cdot{\bf k})^2 + g_{\hbox{eff}}^2 M^2 }$, corresponding to the dispersion of an  electron with wave-vector ${\bf k}$ with spin up or down  ($\sigma = +$ or $-$) relative to the helimagnetic background. A similar replacement is made in the mean field contribution, Eq.(\ref{MeanFieldFreeEnergy}) and $f^\sigma_{\bf k}$ is the Fermi distribution function for occupation of this mode. We have also made a one-loop renormalization of the interaction 
$g_{\hbox{eff}}=g-2g^2 \sum'_{{\bf k}_3,{\bf k}_4} \frac{1}{\epsilon^+_{{\bf k}_1}+\epsilon^-_{{\bf k}_2}
-\epsilon^+_{{\bf k}_3}-\epsilon^-_{{\bf k}_4}}$ 
allowing for the leading order correction to the electron pair wavefunction. 
%
%

\begin{figure}
\includegraphics[width=0.3\textwidth]{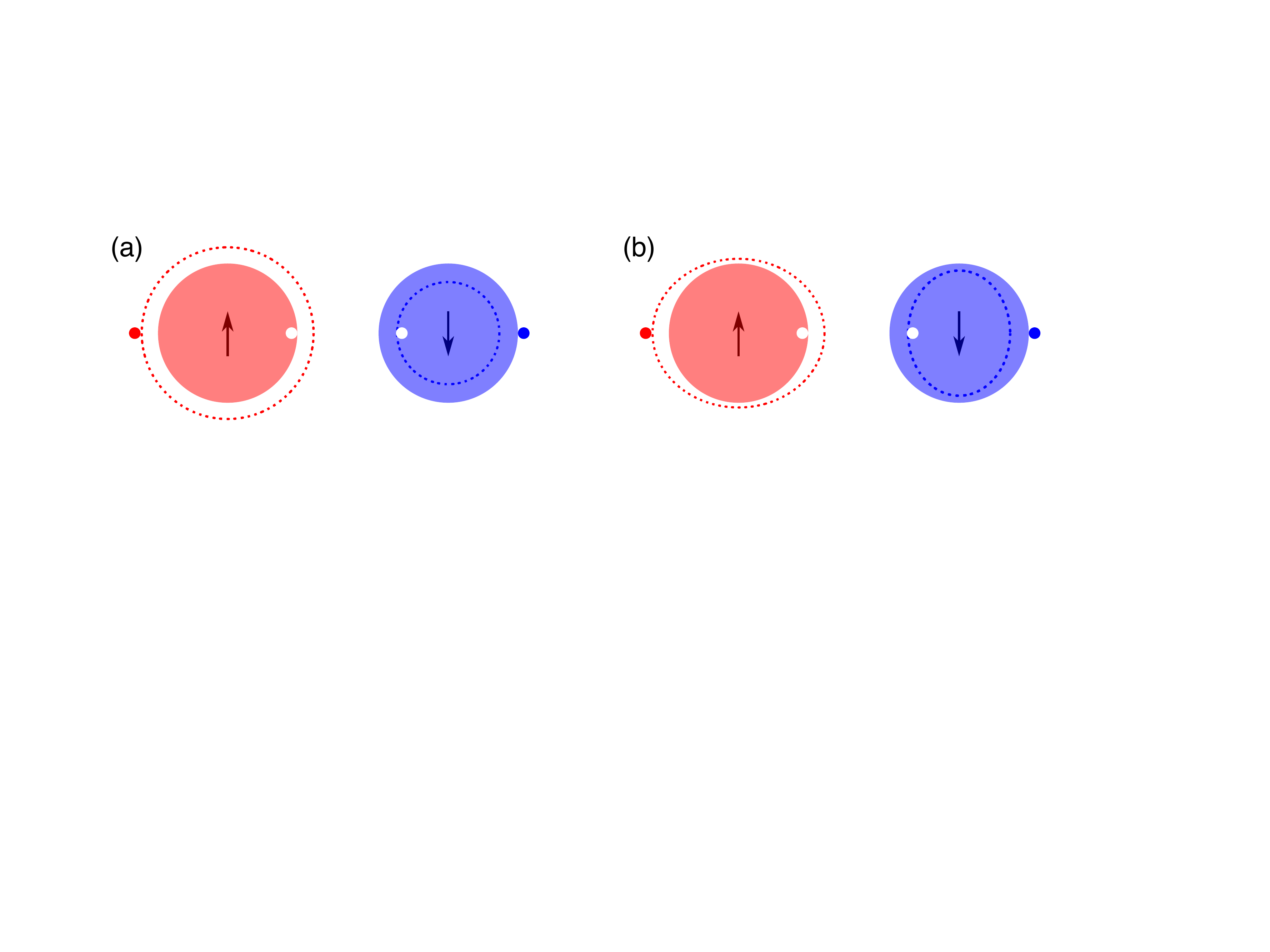}\\
\includegraphics[width=0.3\textwidth]{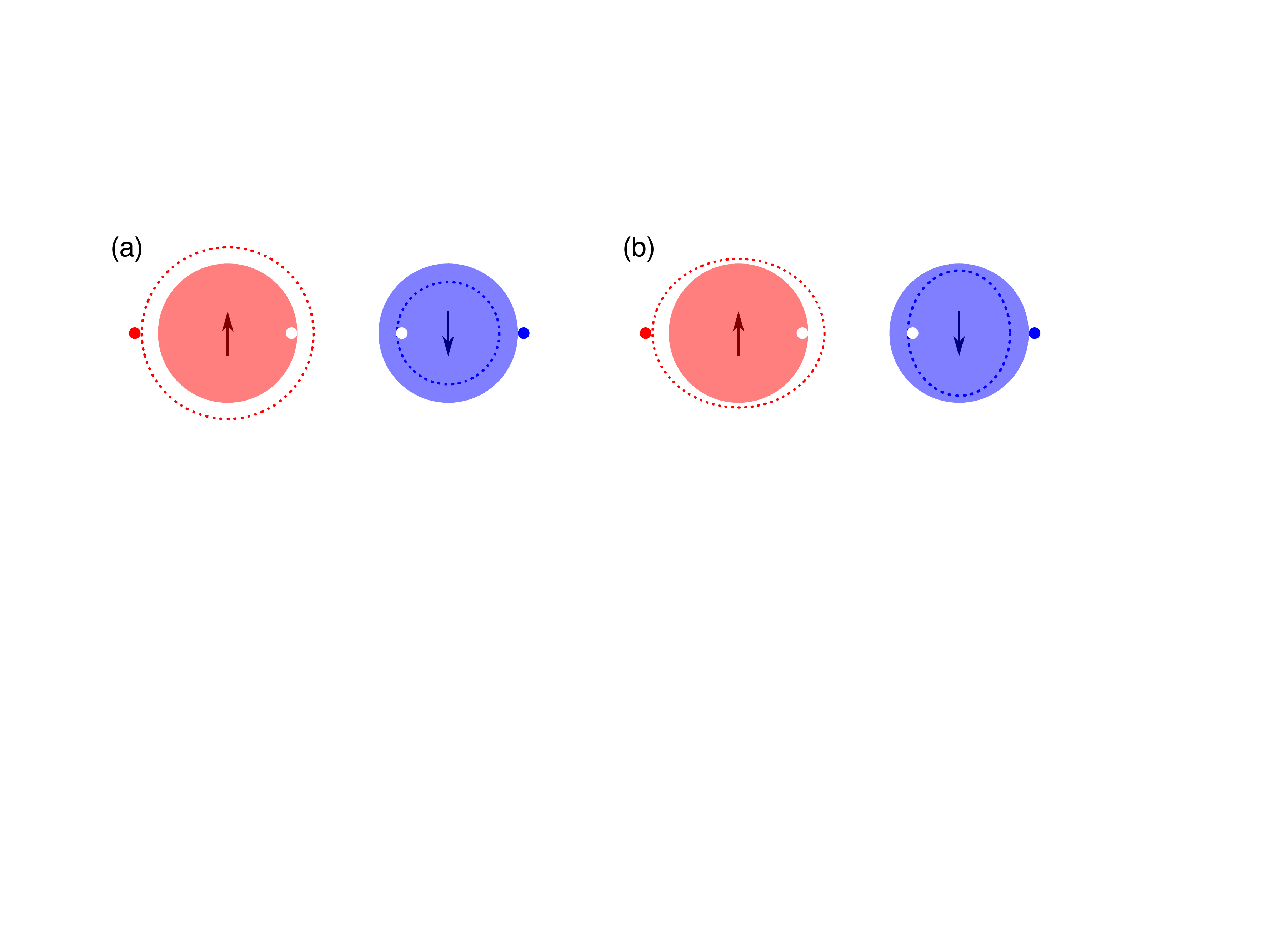}
\caption{
{\it Enhanced low-energy phase space of particle-hole excitations:} (a) Ferromagnetic or (b) helimagnetic distortions of the Fermi surface enhance the low-energy phase space of particle-hole excitations lowering the zero-point energy and driving the transition to ferromagnetic order first order or driving helimagnetic order in the vicinity of the quantum critical point. }
\label{fig:phPhasespace}
\end{figure}

The presence of the background helimagnetic order, $M_{\bf q}$, modifies the mean-field dispersion of the electrons and hence the fluctuation corrections. These fluctuations turn the ferromagnetic transition first order\cite{conduit2009itinerant} and ultimately drive helimagnetic order\cite{conduit09} at sufficiently low temperatures. This can be understood as follows: the second order correction to the energy involves the virtual excitation of pairs of particle-hole pairs --- one spin-up particle-hole pair and one spin-down particle-hole pair --- with total momentum zero. It turns out that the integral in Eq.(\ref{FluctuationCorrection}) is dominated by particle-hole pairs with momentum near $2 k_F$. The density of such excitations at low energy is enhanced by the presence of ferromagnetic or helimagnetic order (see Fig. \ref{fig:phPhasespace}). Since the second order perturbative correction is negative, this lowers the free energy and drives ferromagnetic or helimagnetic order in the region of the phase diagram where fluctuations contributions to the free energy dominate. Indeed, if 
Eq.(\ref{FluctuationCorrection}) is expanded in powers of $M$ and ${\bf q}$ one finds\footnote{
The first calculation of the integrals in Eq.(\ref{FluctuationCorrection}) was carried out in Refs.\cite{huang1957quantum,lee1957many,abrikosov1958concerning} at zero magnetic field. To date, no closed form for the integral at finite $M$ has been found. Initial studies\cite{conduit2009itinerant,duine2005itinerant} calculated the integral numerically and demonstrated how fluctuations could drive the transition in to the ferromagnet first order. Later, leading singularities where calculated analytically and re-summed\cite{pedder2013resummation}.
}
 that the leading singular terms $M^4$ and $|{\bf q}|^2 M^2$ both have coefficients proportional to $\log T$ \cite{conduit09}. This reflects the singularities found in diagrammatic analyses\cite{belitz1997nonanalytic,betouras2005thermodynamics,rech2006quantum,efremov2008nonanalytic,maslov2009nonanalytic}. At sufficiently low temperatures near the critical point, these effects win out and drive first order transitions and helimagnetic order. As temperature is lowered below this point, singularities at higher and higher order in $M$ and $|{\bf q}|$ dominate\cite{pedder2013resummation}. 
At first glance, Eq.(\ref{FluctuationCorrection}) appears to be a term in an expansion in the interaction strength. Obviously this is not the case as the free energy is a functional of the mean-field electron dispersion, which contains the interaction through its dependence upon $M$. In fact, the expansion is in $e^{- k_F \xi}$ where $\xi $ is a typical lengthscale of the interaction potential. As shown by Conduit and Keyserlink in Ref.\cite{Keyserlingk13}, the fluctuation correction Eq.(\ref{FluctuationCorrection}) contains a factor in its integrand  given by the square of the Fourier transform of the interaction potential, $|V({\bf  k}_1-{\bf k}_3)|^2$. 
Since the integrals in Eq.(\ref{FluctuationCorrection}) are dominated by $|{\bf  k}_1-{\bf k}_3| \approx 2 k_F$, these factors are proportional to $e^{- k_F \xi}$ where $\xi $ is a typical lengthscale of the potential. The mean field dispersion does not contain this exponential suppression of the interaction since the potential enters there with near zero momentum.

\subsubsection{Path Integral Formulation}
A formal derivation of Eqs.(\ref{MeanFieldFreeEnergy},\ref{FluctuationCorrection}) from a path integral was presented in Ref.\cite{karahasanovic2012quantum}. Whilst the full details are not important here, aspects are of note because of the connection that they reveal to the Moriya-Hertz-Millis theory, the approximations used and their limitations. The derivation of  proceeds {\it via} the following steps:  First, a fermionic path integral is constructed for the partition function, and the contact interaction decoupled in both spin and charge channels using a Hubbard-Stratonovich transformation. 
At this stage, the Moriya-Hertz-Millis approach (which decouples just the spin part of this interaction) would be to integrate out the electrons in favour of an effective theory for the Hubbard-Stratonovich spin field. Quantum order-by-disorder, makes a simple, but important modification to this. Anticipating the possibility of static magnetic or charge order,  the zero- and finite-frequency parts of the Hubbard-Stratonovich fields are separated. The electrons (whose action is now quadratic) are then integrated out and the resulting effective action for the finite-frequency spin and charge fluctuations in the static  background is truncated to quadratic order. Finally, the finite-frequency charge and spin fluctuations are integrated out to obtain a Ginsburg-Landau function for the spin field\footnote{And potentially the charge field, although in applications to date, it has been assumed that there is no static spatial modulation of the charge field, which can then be absorbed into the chemical potential}. 

By organising the calculation in this way, the comparison with Villain's order-by-disorder\cite{villain1980order} is transparent. 
Villain considered magnetic fluctuations about some mean-field classical background in an approximation that treated them as non-interacting. The resulting bosonic Hamiltonian could be diagonalised by a Bogoliubov transformation from which the zero-point energy  (and entropic contribution to the free energy) was deduced. Here, by separating zero- and finite-frequency parts of the Hubbard-Stratonovich fields, the propagation of electrons is explicitly calculated in the presence of some static, background order, thus accounting (after integrating out the electrons) for the modification of spin and charge fluctuations that results. Expansion of the effective action of the finite-frequency fluctuations to quadratic order, similarly mirrors the non-interacting approximation. 

Moreover, the following argument demonstrates that there are appreciable zero-point fluctuations: Within the non-interacting approximation, the operator 
$\hat b^\dagger_{{\bf q},(\sigma,{\bf p})} = \hat c^\dagger_{{\bf p}+{\bf q}\sigma} \hat c_{{\bf p}\sigma}$ with $|{\bf p}+{\bf q}| > k_F$ and $|{\bf p}|< k_F$ can be considered a bosonic creation operator for spin-$\sigma$ particle-hole excitations at momentum ${\bf q}$ and additional internal index ${\bf p}$. The mean-field Fermi surface provides a vacuum for these excitations.  The propagation of these composite bosonic particles is not trivial --- it is accounted for by precisely the electronic polarization loops that generate Landau damping. However, the interaction part of the Hamiltonian can be expressed simply and contains anomalous terms of the form
$ g \sum_{\bf q} \sum'_{{\bf p},{\bf k}}
\left[
b^\dagger_{-{\bf q},(\uparrow, {\bf p})} 
b^\dagger_{{\bf q},(\downarrow,{\bf k})}
+
b_{{\bf q},(\uparrow, {\bf p})} 
b_{-{\bf q},(\downarrow, {\bf k})} 
\right]$,
where the prime indicates the appropriately restricted summations over the momenta ${\bf p}$ and ${\bf k}$. It is evident that a Bogoliubov transformation is required to diagonalize the bosonic Hamiltonian, suggesting significant zero-point fluctuations and a groundstate of the form\footnote{Compare, for example to the dressing of the Ne\'el state by spinwave excitations: 
\unexpanded{
$$
| \psi \rangle
=\exp \left[ \sum_{{\bf k}} (u_{{\bf k}}/v_{{\bf k}}) \hat b^\dagger_{{\bf k}} \hat b^\dagger_{-{{\bf k}}} \right] 
| ... \uparrow \downarrow \uparrow \downarrow ... \rangle,
$$
}
where $u_{{\bf k}}$ and $v_{{\bf k}}$ are the coefficients used in the Bogoliubov diagonalisation of the 
spinwave Hamiltonian.}
$$
|\psi \rangle = \exp
\left( \sum_{{\bf q}} \sum'_{{\bf p},{\bf k}} U_{{\bf q},{\bf k},{\bf p} } b^\dagger_{-{\bf q},(\uparrow, {\bf p})} 
b^\dagger_{{\bf q},(\downarrow,{\bf k})}
\right)
| MF\rangle
$$
{\it i.e.} dressing the Fermi surface with pairs of spin-up and spin-down particle-hole pairs.

\subsection{Phase Reconstruction}

The singular contribution of zero-point fluctuations to the free energy of the ferromagnet drives a rich variety of phases, even in the simple model of Eq.(\ref{TheHamiltonian}). As indicated in Fig. \ref{fig:PhaseDiagram}, these include a first order transition out of the paramagnetic phase at low temperatures, helimagnetic order, the coexistence with p-wave superconductivity, and also the possibility of spin-antisymmetric nematic order. Next,
we will discuss how all of these can be accommodated within a fermionic order-by-disorder treatment. Including additional, experimentally relevant terms in the Hamiltonian can drive further effects, which we will turn to later.

\subsubsection{1st Order transition and Helimagnetism}

{\it 1st Order Transition:}
Fluctuations generate a diverging negative contribution to the $M^4$-coefficient, which overcomes the positive mean-field contribution and drives the transition first order at a finite-temperature tri-critical point (See Fig.\ref{fig:PhaseDiagram}), as found diagrammatically\cite{belitz1997nonanalytic}.
As the phase boundary is followed to lower and lower temperatures, its position is determined by singularities at higher and higher powers of $M$. In order to track the transition faithfully down to zero-temperature, these increasingly singular contributions must be re-summed  as in Ref.\cite{pedder2013resummation}. Note that the form $M^4 \log M$ at zero-temperature is non-analytical in $M$ which invalidates a conventional Ginzburg-Landau expansion. It does not however invalidate the self-consistent calculation of fluctuations at finite magnetisation employed in fermionic  quantum order-by-disorder.  

{\it Helimagnetism:}
The fluctuation contribution to the free energy Eq.(\ref{FluctuationCorrection}) also contains the seeds of helimagnetic order. Heuristically, the ferromagnetic distortion increases the low-energy phase space for pairs of particle-hole excitations with momentum near $2k_F$ and opposite spin. In fact, any spin anti-symmetric distortion will also enhance this low-energy phase space and be favoured  by the resulting lowering of zero-point energy (see Fig.\ref{fig:phPhasespace}).

These arguments can be made quantitative by noting that the free energy is a functional of the mean-field electron dispersion, $F \equiv F[\epsilon_{\bf k}] $. Aside from a classical contribution $\int d^3 {\bf x} M^2$, the dependence upon the order parameter comes entirely from this dependence. The mean-field electron dispersion in the presence of a helimagnetic order ${\bf M}_{\bf q}=M (\cos {\bf q} \cdot {\bf x} , \sin {\bf q} \cdot {\bf x} ,0)$ is given --- for a quadratic bare dispersion of electrons --- by
$\epsilon^\sigma_{{\bf k}, M_{\bf q}}=\epsilon_{\bf k} - \sigma \sqrt{ ({\bf k} \cdot {\bf q})^2 + M^2}$, describing the propagation of electrons with momentum ${\bf k}$ and spin parallel or antiparallel ($\sigma =+$  or $-$) to the background helimagnetic order. When the dispersion can be linearised at the Fermi surface, the pitch of the helimagnetism ${\bf q}$ enters as a directionally dependent magnetization on the Fermi surface. This implies that coefficients of $M^{2 \alpha}$ are proportional to those of $ |{\bf q}|^{2 \beta} M^{2(\alpha - \beta)}$. When the coefficient of $M^4$ becomes negative at the tricritical point, the coefficient of $|{\bf q}|^2 M^2$ also becomes negative indicating a minimum of the free energy at finite wave-vector and the formation of magnetisation at that wave-vector. 

\subsubsection{Superconductivity and Nematic Order}

The analysis of how fluctuations drive the ferromagnetic transition first order was helped by the fact that the interaction had finite weight in the ferromagnetic channel. Instabilities such as superconductivity or spin anti-symmetric Pomaranchuk/nematic order do not have finite weight in the bare contact interaction considered in Eq.(\ref{TheHamiltonian}). 
It is well-understood diagrammatically how interactions may be generated in these channels\cite{anderson1973anisotropic}.  An appealing picture of how spin fluctuations might drive p-wave superconductivity in a metallic ferromagnet was given in Ref.\cite{julian2012viewpoint}: an electron induces a polarization cloud in the electronic fluid. The net energy of the polarization around two parallel spins is reduced when the clouds overlap. In essence this describes a reduction in zero-point energy upon the formation of Cooper pairs.
One must think a little to recover the same effects in the fermionic quantum  order-by-disorder approach\cite{conduit2013fluctuation}.
The central idea is to calculate a generating function for the new type of order. For example, in the case of p-wave superconductivity in the ferromagnet, a term 
$\sum_{\bf k} 
\left( j_\Delta \theta_{\bf k} c^\dagger_{{\bf k},\uparrow} c^\dagger_{-{\bf k},\uparrow} + c.c \right)$
is added to the Hamiltonian, where $j_\Delta$ is a source field for the superconducting order parameter, $\Delta= \sum_{\bf k} \langle \theta_{\bf k} c^\dagger_{{\bf k},\uparrow} c^\dagger_{-{\bf k},\uparrow} \rangle$. The analysis then proceeds as before, diagonalising the quadratic parts of Hamiltonian and calculating the second order perturbation to the free energy.
Diagonalisation of the Hamiltonian in the presence of these sources requires a Bogoliubov transformation, which modifies the interaction vertex. It is this modification that accounts for the generation of new interaction channels found in the diagrammatic analysis. 
 Once the fluctuation corrections to the free energy have been calculated, a Legendre transformation gives the Ginzburg Landau function. A similar approach has been used to study spin-antisymmetric Pomeranchuk or nematic distortions of the Fermi surface\cite{karahasanovic2012quantum}.

A trick\cite{conduit2013fluctuation} may be used to calculate the Ginzburg-Landau function for continuous transitions: Adding and subtracting a term $\sum_{\bf k} 
\left( \Delta \theta_{\bf k} c^\dagger_{{\bf k},\uparrow} c^\dagger_{-{\bf k},\uparrow} + c.c \right)$ 
to the Hamiltonian. Diagonalising the quadratic parts of the Hamiltonian alongside the added $\Delta c^\dagger c^\dagger$-term and treating the remaining terms alongside the subtracted $\Delta c^\dagger c^\dagger$-term perturbatively, recovers the result of adding a source followed by Legendre transformation up to quadratic order. The saddle-point equations that result from this analysis up to quadratic order are precisely the Eliashberg equations found in diagrammatic, spin-fluctuations calculations.

\subsubsection{Commentary}
Fermionic quantum order-by-disorder provides a complementary formulation of the physics embodied by diagrammatic analyses. Its alternative perspective unifies the particle-instabilities (such as first order transitions and the formation of helimagnetism) of the itinerant ferromagnet, and particle-hole instabilities such as superconductivity. In conventional, diagrammatic analyses, these effects are treated very differently, the former through non-analyticities of the Moriya-Hertz-Millis free energy, and the latter through spin-fluctuation generated pairing in the appropriate channel. The discussion above shows how these effects are borne out in a simple model. 

The approach also provides more direct access to new phenomena. Just as in the case of classical, entropically-driven phase transitions, focussing upon the free energy of fluctuations directly can be very useful in deducing the possibility of other phases and instabilities before a detailed calculation has been performed. Knowing the spectrum of fluctuations and how they are affected by the introduction of certain order can be sufficient to predict the formation of new phases --- even  in the absence of a microscopic model. The formulation has the advantage that calculations follow directly from it. In the following section, we outline a number of developments to fermionic quantum order-by-disorder driven by unusual experimental observation.

\section{EXPERIMENTALLY DRIVEN DEVELOPMENTS}

The experimental systems to which fermionic quantum order-by-disorder can be applied fall into two main classes, ultra-cold atomic gasses and solid-state itinerant magnets. Experimental capabilities for these two types of system are very different, placing consequently different demands upon theoretical models.

\subsection{Ultra-cold Atomic Gases}

Ultra-cold atomic gases offer a pristine arena in which to study collective quantum phenomena.  Interactions can be controlled with great precision by tuning Feshbach resonances, requiring accurate predictions from theory. Amongst the first applications of fermionic quantum order-by-disorder was to understand the ferromagnetic instability of cold fermionic atoms. Early predictions of the scattering length at which this would occur were crucially important\cite{duine2005itinerant,conduit2009repulsive}. Moreover, since atomic gases are naturally in the canonical ensemble, without a mechanism to change the total spin of the gas, it is natural to consider systems with a fixed, spin imbalance\cite{conduit2009itinerant,Parish:2007bh}.

Of course, the study of atomic gases is not without its experimental difficulties. When tuned to positive scattering lengths, there is an unavoidable loss process\cite{conduit2011effect,Pekker11}, which affected the  first attempt to observe ferromagnetic behavior~\cite{Jo09}.
Subsequent experiments have focussed on the
polaron limit~\cite{Kohstall12} for which variational wavefunctions may be presented\cite{bugnion2013ferromagnetic}. Going forward the cold atom gas offers
a chance to explore not only ferromagnetism with the contact
interaction, but also systems with interactions with an effective
range~\cite{Keyserlingk13}, a two-dimensional gas~\cite{conduit13}, and
a mass imbalance between up and down-spin particles~\cite{Keyserlingk11}.

\subsection{Metallic Ferromagnets}

\begin{widetext}

\begin{figure}
\includegraphics[width=0.95\textwidth]{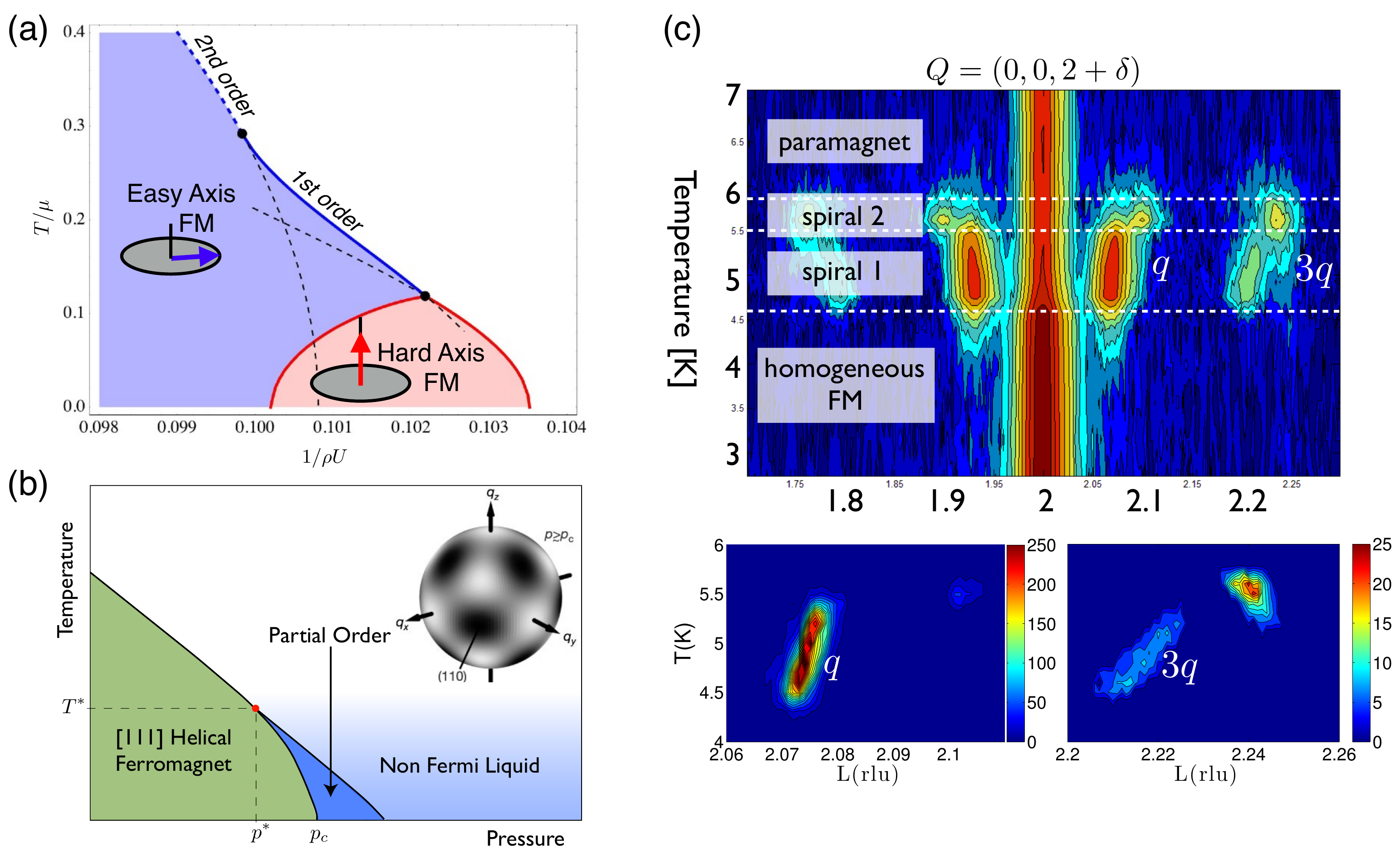}
\caption{
Three experimental situations where the fermionic quantum order-by-disorder approach has proven fruitful,
{\it (a) Hard axis magnetic order:} In materials such as YbRh$_2$Si$_2$ and Yb Ni$_4$P$_2$, hard axis magnetic susceptibility in the paramagnetic phase gives way to ferromagnetic order along this direction in the ordered phase. Fermionic quantum order-by-disorder captures this by an inversion of the magnetic anisotropic in the zero-point fluctuation contributions to the free energy. The figure indicates the phase diagram calculated for this effect in a simple itinerant model.
{\it (b) Partial Order in MnSi:} A variety of explanations have been posited for the partial order in MnSi, including the incipient formation of Skyrmion lattices. Fermionic quantum order-by-disorder provides a relatively simple explanation; orienting the helimagnetic order in directions disfavoured in mean field (away from $[111]$) lowers the energy of fluctuations and is thus favoured when fluctuations are strong near the quantum critical point. The phase diagram indicates that this partial ordered phase occurs in the same region of the phase diagram as the helimagnetic phase of the simple model for the ferromanget.
{\it (c) Helimagnetic Order in PrPtPl:} This material is one of the best examples so far for a fluctuation-induced helimagnetic phase. The figure summarises neutron (top) and x-ray (bottom) scattering data obtained from PrPtAl. This supports the notion that helimagnetic order appears in region of the phase diagram where quantum fluctuations are strong. The theory must be modified to incorporate the effects of local spins. After doing so, a range of thermodynamic and scattering measurements are accounted for by the model.
}
\label{fig:Combined}
\end{figure}
\end{widetext}

Unlike ultra-cold atomic gases, for which precise quantitative predictions from realistic microscopic models are often desirable, in the solid-state we often require minimal models that capture the essence of new emergent phenomena. Fermionic quantum order-by-disorder has provided a good guide in a variety of cases. Ferromagnetic transitions are thought to be generically first-order at low temperature for precisely the reasons illuminated above and this behaviour has been seen in a wide range of materials including Sr$_{1-x}$Ca$_x$RuO$_3$ \cite{uemura2007phase}, CoO$_2$ \cite{Otero-Leal:2008fw}, UGe$_2$\cite{Taufour:2010vk} and URhGe \cite{Yelland:2011xg}. Other cases in which fermionic quantum order-by-disorder has proved illuminating include:

{\it Hard Axis Magnetism:} In a few itinerant metals, such as YbRh$_2$Si$_2$ and YbNi$_4$P$_2$, hard axis magnetic susceptibility in the paramagnetic phase gives way to ferromagnetism along the hard direction in the ordered phase\cite{PhysRevLett.110.256402,steppke2013ferromagnetic,ishida2002b,klingner2011evolution}\footnote{The susceptibilities along the different primary axes cross just above the magnetic transition temperature, the hard axis becoming an easy axis.}. Whilst one can fit such behaviour with an insultating model of frustrated magnetism\cite{andrade2014competing} it is more natural to understand it from the point of view of quantum order by disorder\cite{kruger2014fluctuation}. Establishing ferromagnetic order in the direction anti-favoured at the mean field level costs mean field energy, but leads to a flattened dispersion of low energy excitations and so lower fluctuation contribution to the free energy. When fluctuations dominate, this determines the direction of magnetic order --- {\it i.e.} in the direction anti-favoured by mean field considerations (See Fig.\ref{fig:Combined}). 

{\it Partial order in MnSi:} A similar argument can be used to explain the unusual partially ordered phase of MnSi\cite{Pfleiderer2004partial,kruger2012quantum}. In this material, spin orbit coupling combined with the absence of inversion symmetry conspire to favour helimagnetic order, with a very long pitch
\cite{ishikawa1976helical,bak1980theory,Pfleiderer:2001zv}. Crystalline anisotropy further favours the pitch vectors lining up along the $[111]$ directions. In the partially, ordered phase near to the quantum critical point, neutron scattering shows a smooth distribution of orientations of the pitch vector over the sphere that if anything anti-favours the $[111]$ directions with a slight increase in the pitch of the helimagnetic order. This behaviour arises naturally with a minimal modification of the ferromagnetic Hamiltonian, Eq.(\ref{TheHamiltonian}), that allows for the Dzyalosinksii-Moriya spin orbit coupling and fluctuations (See Fig.\ref{fig:Combined}). Directing the helimagnetism away from the mean-field favoured directions flattens the dispersion of excitations, lowering their zero-point energy. 

{\it Fluctuation-induced helimagnetism:} The experimental search for fluctuation-induced helimagnetism has proven more difficult. Whilst there is circumstantial evidence for it in ZrZn$_2$\cite{uhlarz2004quantum} and NbFe$_2$\cite{crook1995magnetic}, this has not been conclusive. Indeed, it was noted in Ref.\cite{thomson2013helical} that the fluctuation-induced helimagnet might be rather susceptible to the effects of disorder, possibly explaining the apparent spin glass behaviour in the region of the phase diagram of CeFePO(Ref.\cite{Lausberg:2012wq}) where helimagnetism might otherwise be predicted. 
More recently, two materials have emerged as compelling candidates for the emergence of fluctuation-induced helimagnetic order, PrPtAl\cite{abdul2015modulated} and LaCrGe$_3$\cite{lin2013suppression,taufour2016ferromagnetic}. To capture the properties of PrPtAl, the simple model of Eq.(\ref{TheHamiltonian}) must be supplemented with local spins. A range of thermodynamic and scattering measurements are accounted for by the predictions of quantum order-by-disorder (Fig.\ref{fig:Combined}). 

\subsection{Experimental Prospects --- Beyond the Ferromagnet}

\begin{widetext}

\begin{figure}
\includegraphics[width=0.9\textwidth]{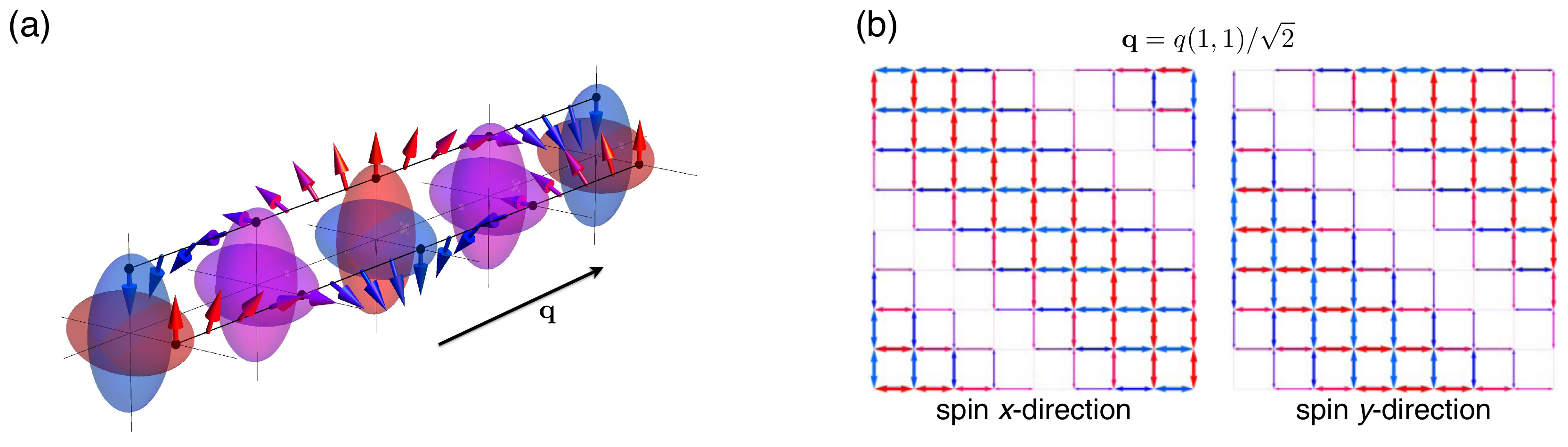}
\caption{
{\it Twisted Spin-Antisymmetric Nematic Order:} Translating the mechanism of fermionic quantum order-by-disorder to the spin-antisymmetric nematic reveals an unusual intertwining of helimagnetic and spin-antisymmetric nematic order whereby the spin quantization axis of the spin-antisymmetric nematic order rotates in space. a) illustrates this order in the continuum showing how electrons with opposite spin have d-wave distortions of their fermi-surfaces with opposite sign. The quantization of this spin axis rotates upon moving in the direction of the pitch vector ${\bf q}$. b) illustrates the projection of this order onto a square lattice, demonstrating its correspondence to a bond order modulation. }
\label{fig:TwistedSpinNematic}
\end{figure}
\end{widetext}

The physics underlying fermionic quantum order-by-disorder is very general and is expected to find application far beyond the itinerant ferromagnet to which it has initially been applied. For example, the {\it spin-triplet nematic state} first proposed in \cite{kivelson1998electronic} and studied in a simple model by a number of authors\cite{
pomeranchuk1958stability,
oganesyan2001quantum,
quintanilla2006pomeranchuk,
kee2003signatures,
khavkine2004formation,
kee2005itinerant,
yamase2005mean,
doh2006interplay,
lawler2006nonperturbative,
wu2007fermi,
lawler2007local} was first shown to have its transition driven first order by fluctuations in Ref.\cite{kirkpatrick2011nature}. It was later shown \cite{hannappel2016stability}, using fermionic quantum order-by-disorder, that in fact fluctuations promote the formation of a new phase near to the quantum critical point that intertwines magnetic modulation and $d$-wave orbital order in a continuum version of bond density wave order(see Fig.\ref{fig:TwistedSpinNematic}). These unusual phases have some intriguing observable consequences. For example, whereas the static spin-triplet
nematic responds to a uniform magnetic field by generating
an anisotropic strain\cite{wu2007fermi}, the triplet d-density wave generates
a spatially modulated strain. This offers new possibilities for
experimentally isolating multipolar order.

The  interplay of fluctuations with band effects is of particular importance in real-life systems. The band dispersion itself may lead to phase reconstruction in some ways similar to that caused by fluctuations driving transitions first order when the Wolfharth-Rhodes criterion is satisfied\cite{wohlfarth1962collective} or leading to spin-antisymmetric nematic or spin density wave due to weak nesting as found in Sr$_3$RuO$_3$\cite{grigera2004disorder,berridge2009inhomogeneous}. A proper account of this interplay has not yet been given, though a preliminary analysis was presented in \cite{GarethsThesis}.
The extreme of such interplay occurs in the {\it critical antiferromagnet} in which spin fluctuations are thought to drive $d$-wave superconductivity\cite{fradkin2013field,zaanen2006towards,nagaosa1999quantum}, and in which the subtle interplay of band effects and spin fluctuations still leaves open questions. Analysis using fermionic quantum order-by-disorder may be useful\cite{HannappelThesis}.

A range of other systems of experimental interest may also be analysed using fermionic quantum order-by-disorder. For example, the lattice Dirac fermions found in graphene harbour a number of instabilities that might be captured in this way\cite{
lemonik2012competing,
dahal2010charge,
scherer2012instabilities}.
Similarly, the gradual addition of new features, such as spin-orbit coupling, topological bandstructures {\it etc.}, to the toy Hamiltonian, Eq.(\ref{TheHamiltonian}) might reveal new and unusual features. The possibilities are very rich and best explored in concert with experiment. 

\section{RELATION TO OTHER APPROACHES}

As we have emphasised, fermionic quantum order-by-disorder provides a complementary view on the role of fluctuations in forming exotic quantum phases. We have emphasised the relation to spin-fluctuation theories and --- in the application to the itinerant ferromagnet --- non-analyitic extensions of Moriya-Hertz-Millis theory. There are a variety of other approaches to which it is related in spirit, and it is worth mentioning some of them here. 

\subsection{QOBD vs DFT, DMFT and Kadanoff-Baym}

Whilst analytically tractable models play an important role in the qualitative understanding of new phenomena, sometimes numerically precise predictions are needed, particularly when guiding experiment through a delicate balance of competing possibilities. Though progress has been made towards first principle calculations\cite{essenberger2014superconducting,essenberger2016ab}, presently, the calculation of fluctuation-induced effects, such as superconductivity, is carried out in parallel with {\it ab initio} band structure calculations; the latter providing spin susceptibilities that are used in field theoretical calculations. It would be appealing if these calculations could be performed in concert. QOBD suggests a natural way to do so. It sits naturally in the framework described in Ref.\cite{kotliar2006electronic}, which places density functional theory, dynamical mean field and Baym-Kadanoff theory in the unified context of Legendre transformation of an appropriate generating functional. The potential for systematically  including these effects into {\it ab intio} codes has great promise. 

Fermionic quantum order-by-disorder is essentially a restriction of Baym-Kadanoff theory where the variational parameter --- the Green's function --- is itself characterised by a restricted set of parameters given by the order parameters of the fields that we wish to study. Indeed, the additional fluctuation corrections resulting from the inclusion of superconducting order (see Ref.\cite{conduit2013fluctuation}  where these results are obtained as outlined in Section III C), are essentially the modfications to the density functional used for {\it ab initio} calculations of electronically mediated superconductivity in Refs.\cite{essenberger2014superconducting,essenberger2016ab}.
%
%
 This presents broader possibilities for the inclusion of novel order, such as spin-antisymmetric nematic order,  in density functional code. If successful, this would complement the analytical description of minimal models with numerical accuracy when applied to real materials, and would be a potentially fruitful avenue of future study.

\subsection{Functional Renormalisation Group }
Fluctuations-driven ordering phenomena are naturally captured in renormalisation-group (RG) approaches. As discussed previously, certain instabilities are not directly supported by short range Coulomb repulsion, but fluctuations renormalise the low-energy theory and generate (at one-loop order) new interaction vertices that support the instability. The $p$-wave superconductivity in metallic ferromagnets\cite{anderson1973anisotropic} is a good example. A more recent example are the spin-fluctuation driven nematic and superconducting instabilities of iron-based superconductors \cite{Fang+08,Fernandes+12}. That the same physics can be described by the QOBD mechanism \cite{conduit2013fluctuation} shows the close relation between the two approaches. 

The functional renormalization group (fRG) is, in principle, an exact method that can be applied to both bosonic \cite{Wetterich93} and fermionic \cite{Metzner+12} systems. A comprehensive summary can be found in the review by Salmhofer \cite{Salmhofer99}. Instead of the scale dependence of a finite number of coupling constants, fRG keeps track of the full momentum and frequency dependence of the single-particle Green function and interaction vertices. It is  a flexible and unbiased tool to study scale dependent behavior in electron systems, e.g. competing magnetic, charge, and pairing instabilities and the interplay between electronic excitations and order-parameter fluctuations. One of the main advantages of the fRG-based one-loop computation of the two-particle vertex, 
compared to other weak-coupling approaches, is that particle-particle and particle-hole channels are treated on an equal footing, without artificial bias towards a particular channel of instability \cite{Metzner+12}. 

Practical implementations require approximations. Since the flow of a given vertex depends upon higher-order vertices, the effective action must be truncated at some order. Moreover, an infrared cut-off $\Lambda$ that serves as RG flow parameter needs to be introduced in the free-fermion action. Typical choices are a momentum cut-off that defines a small region around the Fermi surface, similar to the Wilsonian RG approach to interacting fermions \cite{Benfatto+90,Shankar91,Shankar94,Polchinski92}, or a frequency cut-off. Once self-energy effects are taken into account, the Fermi surface is usually deformed in the course of the RG flow.  Using a frequency cut-off has the advantage that it does not interfere with Fermi-surface deformations and that particle-processes with small momentum transfer are captured smoothly by the flow \cite{Husemann+09}. 
Finally, numerical integration requires a discretisation of the fRG equations --- usually achieved by introducing a mesh in momentum space. Because of this, subtle changes 
associated with the onset of an exotic order parameter could be missed. For example, the fluctuation-driven helical magnetic order close to ferromagnetic quantum-critical points has a very large periodicity, corresponding to very small  Fermi-surface deformations. 

Because of the computational complexity, fRG has so far only been successfully applied to one- and two-dimensional electron systems. In this context, it is interesting to mention that early fRG studies of graphene-type systems predicted a topological quantum spin-Hall state for strong next-nearest neighbor interactions \cite{Raghu+08}. More recent fRG studies with a much higher momentum resolution, however, show that this transition is pre-empted by the formation of a three-sublattice, charge-modulated state \cite{Pena+17}. It would be interesting to determine the predictions that QOBD makes for this system.
While the fRG equations in their full form are exact, the commonly used truncation schemes restrict their applications to the study of weak-coupling instabilities. At present it remains an open question which truncation schemes of the fRG equations are required to study the full range of fluctuation-driven ordering phenomena captured by the QOBD mechanism. 

\subsection{Monte Carlo}

The exquisite precision of experiments on ultra-cold gases alluded to in Section IV. A, places exacting demands upon theoretical predictions. Precise determination of scattering lengths (the preferred measure of interaction strength) at which phase transitions and instabilities occur must be met with similarly precise calculations. 
Various quantum Monte-Carlo calculations have been performed to achieve this --- in broad agreement with QOBD. 

{\it Quantum Monte Carlo}~\cite{Needs2010} provides an accurate numerical method
to probe the effects of quantum fluctuations in strongly correlated systems. 
The method not only offers a complementary analysis underpinned by different
approximations to the QOBD formalism, but also a precise level of
control over the Hamiltonian and in particular the interaction potential.
The starting point for fermionic systems is a Slater-Jastrow trial
wavefunction~\cite{Needs2010}, $\psi=e^{J}D_{\uparrow}D_{\downarrow}$.
Here $D_{\{\uparrow,\downarrow\}}$ is a Slater determinant of the
up-/down-spin particle plane-wave states corresponding to the
non-interacting ground state, and ensures the correct fermion
antisymmetry. Different polarizations are addressed by changing
the number of up-/down-spin particles in their respective determinants.
$J$ is a Jastrow factor that contains additional variational terms in
particle-particle separation~\cite{Whitehead16},
$J=\sum_{n,i,j}\alpha_{n,\sigma_{i},\sigma_{j}}(r_{i}-r_{j})^n$. 
In Variational Monte Carlo the parameters
$\{\alpha_{n,\sigma_{i}\sigma_{j}}\}$ are optimized to minimize the
ground state energy. Further refinement of this wavefunction is can be 
obtained using Diffusion Monte Carlo; an accurate
Green's function method that projects out the ground state~\cite{Needs2010}.

 Most ultra-cold gases have contact interactions with zero effective range. These are characterised by the scattering phase shift $\delta$, which can be related to the scattering length $a\ge0$ according to $\cot\delta(k)=-1/(ka)$ (for a zero-range interaction). Several different approaches have been used to capture such interactions in Monte-Carlo:
 \noindent
i. An explicitly repulsive top hat potential~\cite{conduit09}. The
strength and radius of the top hat can be adjusted to give zero
effective range at the cost of introducing higher order terms in the
scattering phase shift~\cite{conduit13}.
\noindent
ii.
The contact interaction limit can be reached by using an
attractive interaction potential with short
range~\cite{Pilati10,Chang11}. However, this potential
harbors a bound state. To avoid the system entering the bound state the
trial wave function must be restricted to have no variational
parameters, meaning that the ground state cannot be reached.
\noindent
iii. The short range behavior of the wave function induced by the
contact interaction can be directly imprinted upon the Jastrow
factor~\cite{Pilati10} 
However, this gives incorrect scattering
properties in the $\ell\ge1$ scattering channels.
Notwithstanding these subtleties, it has proven possible to extract a general,
transferable pseudopotential that gives the correct scattering properties, and which does not harbor a bound state, allowing it to be used in a variationally~\cite{Bugnion14}.

The Quantum Monte Carlo approaches have been used to confirm the
presence of a first order transition
and helimagnetic phase~\cite{conduit09} 
Moreover, 
 these results have also been used to define a density functional\cite{ma2012density}, thus fulfilling part of the aim of bringing QOBD effects into the realm of density functional theory. 
These results provide independent verification of the
QOBD approach: they include all possible correlations, and deliver a
ground state restricted only by the fixed node approximation. 
The Quantum Monte Carlo approach also offers access to two important
extensions: the consequences of effective range
interactions~\cite{Keyserlingk13}, two-dimensional
geometry~\cite{conduit13}, and different masses for the spin-up and spin-down particles~\cite{Keyserlingk11}.
However, the numerical calculations are
performed in a finite sized system and so cannot capture all long wavelength
fluctuations.

\section{BROADER PERSPECTIVE AND CONCLUSIONS}

Fermionic QOBD has had a number of successes --- particularly in its application to ferromagnetic metals. Development of the method to improve its range and accuracy of application suggests several near- and medium-term goals. 

{\it Near-term goals} might include considering the formation of phases near the quantum critical points of different background order --- we discussed above extensions of the original application to ferromagnetic order  to spin-anti-symmetric nematic order and anti-ferromagnetic order. Charge density wave order is another interesting possibility. However, the possibilities for QOBD effects to operate in bands with non-trivial topological order is perhaps more intriguing. Implicit in the latter is the effect of more intricate band structure and spin-orbit coupling. Conventional band structure can itself generate instabilities towards different types of order through nesting and other features in the density of states. A systematic way to treat these alongside fluctuations does not yet exist. Addition of  more ingredients in this way makes assessment of which effects win out more and more delicate --- a systematic, numerical way to accommodate this (for example by inclusion in density functional theory) alongside analytical calculations for minimal toy models is desirable. 

Beyond these considerations, the inclusion of new types of instabilities and exotic order presents a number of {\it medium-term goals}. The description of local critical fluctuations\cite{si2001locally} from this perspective and indeed whether Mott physics can be accessed by QOBD raise important questions. There is some hope for the latter given the similarities between QOBD and the description of the Mott physics using dynamical mean field theory. Indeed a rather direct translation of the simple Landau theory developed in Ref.\cite{kotliar2000landau} may indeed be able to re-express the Mott transition in terms of QOBD. 

Unconventional order such as that found in gapless spin-liquid phases presents even greater challenges\cite{lee2006doping}. The fermionic version of QOBD might possibly be applied to understand instabilities of the spinon-fermi surface. In its wider application there have been recent hints that QOBD might be useful to understand deconfined quantum criticality in terms of fluctuation-induced transitions in entanglement structure\cite{Senthil:2004qi,kuklov2008deconfined,green2016feynman}.  The $1/N$-effects of fluctuations in anti-ferromagnets\cite{read1989valence} can be reframed as a zero-point energy breaking the degeneracy amongst a set of states with different entanglement structure\cite{green2016feynman}. This brings us full circle to the work of Villain\cite{villain1980order} and shows that his perspective can be used to understand a large range of phenomena. 
This broader perspective is closely aligned to one of the most fascinating developments in fundamental physics; the notion that gravity itself might be an entropic force\cite{maldacena1998adv,maldacena2005illusion,verlinde2011origin} 
emerging from zero-point fluctuations. The idea being that the space-time metric can be associated with entanglement structure, and that this entanglement structure is itself determined by quantum fluctuations.  In the case of fundamental physics, these ideas arise from the duality between gauge theory and string theory. Villain's simple notion of order-by-disorder 
applied to condensed matter systems, may yet provide an alternative route to emergent geometry of entanglement. 

Quantum order-by-disorder is one amongst many perspectives from which to view strongly correlated quantum systems. Each perspective privileges a particular way to proceed. In some cases, QOBD allows a particularly simple, heuristic picture of the 
physical processes promoting certain collective behaviour, one that can very directly be turned to concrete calculation. It is an idea whose ramifications have yet to be fully realised. 

\section*{DISCLOSURE STATEMENT}
The authors are not aware of any affiliations, memberships, funding, or financial holdings that
might be perceived as affecting the objectivity of this review. 

\section*{ACKNOWLEDGMENTS}
We acknowledge funding from the EPSRC and Royal Society

%

\bibliographystyle{ar-style4.bst}
\bibliography{bibliography}

\begin{thebibliography}{130}
\expandafter\ifx\csname natexlab\endcsname\relax\def\natexlab#1{#1}\fi

\bibitem{onsager1949effects}
Onsager L. 1949.
\textit{Annals of the New York Academy of Sciences} 51:627--659

\bibitem{Adams:1998kq}
Adams M, Dogic Z, Keller SL, Fraden S. 1998.
\textit{Nature} 393:349--352

\bibitem{BerezinskiiI}
{Berezinski{\v \i}} VL. 1971.
\textit{Soviet Journal of Experimental and Theoretical Physics} 32:493

\bibitem{BerezinskiiII}
{Berezinski{\v \i}} VL. 1972.
\textit{Soviet Journal of Experimental and Theoretical Physics} 34:610

\bibitem{KTtransition}
Kosterlitz JM, Thouless DJ. 1973.
\textit{Journal of Physics C: Solid State Physics} 6:1181

\bibitem{MeersonInvertedPendulum}
Simons YB, Meerson B. 2009.
\textit{Phys. Rev. E} 80:042102

\bibitem{mackay2003information}
MacKay DJ. 2003.
Information theory, inference and learning algorithms.
Cambridge university press

\bibitem{LondonVdW}
{London} F. 1930.
\textit{Zeitschrift fur Physik} 63:245--279

\bibitem{casimir1948attraction}
Casimir HB. 1948.
On the attraction between two perfectly conducting plates. In
  \textit{Proceedings of the KNAW}, vol.~51

\bibitem{anderson1973anisotropic}
Anderson PW, Brinkman W. 1973.
\textit{Physical Review Letters} 30:1108

\bibitem{brinkman1974spin}
Brinkman W, Serene J, Anderson P. 1974.
\textit{Physical Review A} 10:2386

\bibitem{fay1980coexistence}
Fay D, Appel J. 1980.
\textit{Physical Review B} 22:3173

\bibitem{villain1980order}
Villain J, Bidaux R, Carton JP, Conte R. 1980.
\textit{Journal de Physique} 41:1263--1272

\bibitem{Note1}
The very similar effects of zero-point fluctuations and thermal fluctuations is
  revealed particularly clearly; a factor of $n_{\hbox {B}}+1/2$ for each
  spinwave mode accounts for both zero-point (as $T\rightarrow 0$) and thermal
  (as $T\rightarrow \infty $) occupation.

\bibitem{MoessnerOBD1}
Tchernyshyov O, Moessner R, Sondhi SL. 2002.
\textit{Phys. Rev. Lett.} 88:067203

\bibitem{MoessnerOBD2}
Zhitomirsky ME, Gvozdikova MV, Holdsworth PCW, Moessner R. 2012.
\textit{Phys. Rev. Lett.} 109:077204

\bibitem{ParamekantiOBD}
Mulder A, Ganesh R, Capriotti L, Paramekanti A. 2010.
\textit{Phys. Rev. B} 81:214419

\bibitem{hertz1976quantum}
Hertz JA. 1976.
\textit{Physical Review B} 14:1165

\bibitem{landau1946vibrations}
Landau LD. 1946.
\textit{Zh. Eksp. Teor. Fiz.} 10:25

\bibitem{moriya2012spin}
Moriya T. 2012.
Spin fluctuations in itinerant electron magnetism.
vol.~56.
Springer Science \& Business Media

\bibitem{millis1993effect}
Millis A. 1993.
\textit{Physical Review B} 48:7183

\bibitem{belitz1997nonanalytic}
Belitz D, Kirkpatrick T, Vojta T. 1997.
\textit{Physical Review B} 55:9452

\bibitem{Note2}
In fact, the existence of these non-analyticities had been noted earlier\cite
  {PhysRevB.15.1523} but their physical significance was not understood until
  the work of Belitz-Kirkpatrick-Vojta.

\bibitem{betouras2005thermodynamics}
Betouras J, Efremov D, Chubukov A. 2005.
\textit{Physical Review B} 72:115112

\bibitem{rech2006quantum}
Rech J, Pepin C, Chubukov AV. 2006.
\textit{Physical Review B} 74:195126

\bibitem{efremov2008nonanalytic}
Efremov DV, Betouras JJ, Chubukov A. 2008.
\textit{Physical Review B} 77:220401

\bibitem{maslov2009nonanalytic}
Maslov DL, Chubukov AV. 2009.
\textit{Physical Review B} 79:075112

\bibitem{conduit2009itinerant}
Conduit G, Simons B. 2009{\natexlab{a}}.
\textit{Physical Review A} 79:053606

\bibitem{conduit09}
Conduit G, Green A, Simons B. 2009.
\textit{Physical review letters} 103:207201

\bibitem{Note3}
The first calculation of the integrals in Eq.(\ref {FluctuationCorrection}) was
  carried out in Refs.\cite
  {huang1957quantum,lee1957many,abrikosov1958concerning} at zero magnetic
  field. To date, no closed form for the integral at finite $M$ has been found.
  Initial studies\cite {conduit2009itinerant,duine2005itinerant} calculated the
  integral numerically and demonstrated how fluctuations could drive the
  transition in to the ferromagnet first order. Later, leading singularities
  where calculated analytically and re-summed\cite {pedder2013resummation}.

\bibitem{pedder2013resummation}
Pedder C, Kr{\"u}ger F, Green A. 2013.
\textit{Physical Review B} 88:165109

\bibitem{Keyserlingk13}
von Keyserlingk C, Conduit G. 2013.
\textit{Physical Review B} 87:184424

\bibitem{karahasanovic2012quantum}
Karahasanovic U, Kr{\"u}ger F, Green AG. 2012.
\textit{Physical Review B} 85:165111

\bibitem{Note4}
And potentially the charge field, although in applications to date, it has been
  assumed that there is no static spatial modulation of the charge field, which
  can then be absorbed into the chemical potential

\bibitem{Note5}
Compare, for example to the dressing of the Ne\'el state by spinwave
  excitations: $$ | \psi \rangle =\exp \left [ \sum _{{\bf k}} (u_{{\bf
  k}}/v_{{\bf k}}) \hat b^\dagger _{{\bf k}} \hat b^\dagger _{-{{\bf k}}}
  \right ] | ... \uparrow \downarrow \uparrow \downarrow ... \rangle , $$ where
  $u_{{\protect \bf k}}$ and $v_{{\protect \bf k}}$ are the coefficients used
  in the Bogoliubov diagonalisation of the spinwave Hamiltonian.

\bibitem{julian2012viewpoint}
Julian S. 2012.
\textit{Physics} 5:17

\bibitem{conduit2013fluctuation}
Conduit G, Pedder C, Green A. 2013.
\textit{Physical Review B} 87:121112

\bibitem{duine2005itinerant}
Duine R, MacDonald A. 2005.
\textit{Physical review letters} 95:230403

\bibitem{conduit2009repulsive}
Conduit G, Simons B. 2009{\natexlab{b}}.
\textit{Physical review letters} 103:200403

\bibitem{Parish:2007bh}
Parish MM, Marchetti FM, Lamacraft A, Simons BD. 2007.
\textit{Nat Phys} 3:124--128

\bibitem{conduit2011effect}
Conduit G, Altman E. 2011.
\textit{Physical Review A} 83:043618

\bibitem{Pekker11}
Pekker D, Babadi M, Sensarma R, Zinner N, Pollet L, et~al. 2011.
\textit{Physical review letters} 106:050402

\bibitem{Jo09}
Jo GB, Lee YR, Choi JH, Christensen CA, Kim TH, et~al. 2009.
\textit{Science} 325:1521--1524

\bibitem{Kohstall12}
Kohstall C, Zaccanti M, Jag M, Trenkwalder A, Massignan P, et~al. 2012.
\textit{Nature} 485:615--618

\bibitem{bugnion2013ferromagnetic}
Bugnion P, Conduit G. 2013.
\textit{Physical Review A} 87:060502

\bibitem{conduit13}
Conduit G. 2013.
\textit{Physical Review B} 87:184414

\bibitem{Keyserlingk11}
von Keyserlingk C, Conduit G. 2011.
\textit{Physical Review A} 83:053625

\bibitem{uemura2007phase}
Uemura Y, Goko T, Gat-Malureanu I, Carlo J, Russo P, et~al. 2007.
\textit{Nature Physics} 3:29--35

\bibitem{Otero-Leal:2008fw}
Otero-Leal" M. 2008.
\textit{Physical Review B} 78

\bibitem{Taufour:2010vk}
Taufour" V. 2010.
\textit{Physical Review Letters} 105

\bibitem{Yelland:2011xg}
Yelland EA, Barraclough JM, Wang W, Kamenev KV, Huxley AD. 2011.
\textit{Nat Phys} 7:890--894

\bibitem{PhysRevLett.110.256402}
Lausberg S, Hannaske A, Steppke A, Steinke L, Gruner T, et~al. 2013.
\textit{Phys. Rev. Lett.} 110:256402

\bibitem{steppke2013ferromagnetic}
Steppke A, K{\"u}chler R, Lausberg S, Lengyel E, Steinke L, et~al. 2013.
\textit{Science} 339:933--936

\bibitem{ishida2002b}
Ishida K, Okamoto K, Kawasaki Y, Kitaoka Y, Trovarelli O, et~al. 2002.
\textit{Physical review letters} 89:107202

\bibitem{klingner2011evolution}
Klingner C, Krellner C, Brando M, Geibel C, Steglich F, et~al. 2011.
\textit{Physical Review B} 83:144405

\bibitem{Note6}
The susceptibilities along the different primary axes cross just above the
  magnetic transition temperature, the hard axis becoming an easy axis.

\bibitem{andrade2014competing}
Andrade EC, Brando M, Geibel C, Vojta M. 2014.
\textit{Physical Review B} 90:075138

\bibitem{kruger2014fluctuation}
Kr{\"u}ger F, Pedder C, Green A. 2014.
\textit{Physical review letters} 113:147001

\bibitem{Pfleiderer2004partial}
Pfleiderer C, Reznik D, Pintschovius L, L{\"o}hneysen Hv, Garst M, Rosch A.
  2004.
\textit{Nature} 427:227--231

\bibitem{kruger2012quantum}
Kr{\"u}ger F, Karahasanovic U, Green AG. 2012.
\textit{Physical review letters} 108:067003

\bibitem{ishikawa1976helical}
Ishikawa eY, Tajima K, Bloch D, Roth M. 1976.
\textit{Solid State Communications} 19:525--528

\bibitem{bak1980theory}
Bak P, Jensen MH. 1980.
\textit{Journal of Physics C: Solid State Physics} 13:L881

\bibitem{Pfleiderer:2001zv}
Pfleiderer C, Julian SR, Lonzarich GG. 2001.
\textit{Nature} 414:427--430

\bibitem{uhlarz2004quantum}
Uhlarz M, Pfleiderer C, Hayden S. 2004.
\textit{Physical review letters} 93:256404

\bibitem{crook1995magnetic}
Crook M, Cywinski R. 1995.
\textit{Journal of magnetism and magnetic materials} 140:71--72

\bibitem{thomson2013helical}
Thomson S, Kr{\"u}ger F, Green A. 2013.
\textit{Physical Review B} 87:224203

\bibitem{Lausberg:2012wq}
Lausberg" S. 2012.
\textit{Physical Review Letters} 109

\bibitem{abdul2015modulated}
Abdul-Jabbar G, Sokolov DA, O'neill CD, Stock C, Wermeille D, et~al. 2015.
\textit{Nature Physics} 11:321--327

\bibitem{lin2013suppression}
Lin X, Taufour V, Bud'ko SL, Canfield PC. 2013.
\textit{Physical Review B} 88:094405

\bibitem{taufour2016ferromagnetic}
Taufour V, Kaluarachchi US, Khasanov R, Nguyen MC, Guguchia Z, et~al. 2016.
\textit{Physical Review Letters} 117:037207

\bibitem{kivelson1998electronic}
Kivelson SA, Fradkin E, Emery VJ. 1998.
\textit{Nature} 393:550--553

\bibitem{pomeranchuk1958stability}
Pomeranchuk IY, et~al. 1958.
\textit{Sov. Phys. JETP} 8:361

\bibitem{oganesyan2001quantum}
Oganesyan V, Kivelson SA, Fradkin E. 2001.
\textit{Physical Review B} 64:195109

\bibitem{quintanilla2006pomeranchuk}
Quintanilla J, Schofield A. 2006.
\textit{Physical Review B} 74:115126

\bibitem{kee2003signatures}
Kee HY, Kim EH, Chung CH. 2003.
\textit{Physical Review B} 68:245109

\bibitem{khavkine2004formation}
Khavkine I, Chung CH, Oganesyan V, Kee HY. 2004.
\textit{Physical Review B} 70:155110

\bibitem{kee2005itinerant}
Kee HY, Kim YB. 2005.
\textit{Physical Review B} 71:184402

\bibitem{yamase2005mean}
Yamase H, Oganesyan V, Metzner W. 2005.
\textit{Physical Review B} 72:035114

\bibitem{doh2006interplay}
Doh H, Friedman N, Kee HY. 2006.
\textit{Physical Review B} 73:125117

\bibitem{lawler2006nonperturbative}
Lawler MJ, Barci DG, Fern{\'a}ndez V, Fradkin E, Oxman L. 2006.
\textit{Physical Review B} 73:085101

\bibitem{wu2007fermi}
Wu C, Sun K, Fradkin E, Zhang SC. 2007.
\textit{Physical Review B} 75:115103

\bibitem{lawler2007local}
Lawler MJ, Fradkin E. 2007.
\textit{Physical Review B} 75:033304

\bibitem{kirkpatrick2011nature}
Kirkpatrick T, Belitz D. 2011.
\textit{Physical review letters} 106:105701

\bibitem{hannappel2016stability}
Hannappel G, Pedder C, Kr{\"u}ger F, Green A. 2016.
\textit{Physical Review B} 93:235105

\bibitem{wohlfarth1962collective}
Wohlfarth E, Rhodes P. 1962.
\textit{Philosophical Magazine} 7:1817--1824

\bibitem{grigera2004disorder}
Grigera S, Gegenwart P, Borzi R, Weickert F, Schofield A, et~al. 2004.
\textit{Science} 306:1154--1157

\bibitem{berridge2009inhomogeneous}
Berridge A, Green A, Grigera S, Simons B. 2009.
\textit{Physical review letters} 102:136404

\bibitem{GarethsThesis}
Conduit G. 2009.
Collective phenomena in correlated semiconductors, degenerate fermi gases, and
  ferroelectrics.
Ph.D. thesis, University of Cambridge, Department of Physics

\bibitem{fradkin2013field}
Fradkin E. 2013.
Field theories of condensed matter physics.
Cambridge University Press

\bibitem{zaanen2006towards}
Zaanen J, Chakravarty S, Senthil T, Anderson P, Lee P, et~al. 2006.
\textit{Nature Physics} 2:138--143

\bibitem{nagaosa1999quantum}
Nagaosa N. 1999.
Quantum field theory in strongly correlated electronic systems.
Springer Science \&amp; Business Media

\bibitem{HannappelThesis}
Hannappel G. 2016.
The role of fluctuations near antiferromagnetic and spin-triplet nematic
  quantum critical points.
Ph.D. thesis, University College London, http://discovery.ucl.ac.uk/1489651/

\bibitem{lemonik2012competing}
Lemonik Y, Aleiner I, Fal'ko V. 2012.
\textit{Physical review b} 85:245451

\bibitem{dahal2010charge}
Dahal HP, Wehling TO, Bedell KS, Zhu JX, Balatsky A. 2010.
\textit{Physica B: Condensed Matter} 405:2241--2244

\bibitem{scherer2012instabilities}
Scherer MM, Uebelacker S, Honerkamp C. 2012.
\textit{Physical Review B} 85:235408

\bibitem{essenberger2014superconducting}
Essenberger F, Sanna A, Linscheid A, Tandetzky F, Profeta G, et~al. 2014.
\textit{Physical Review B} 90:214504

\bibitem{essenberger2016ab}
Essenberger F, Sanna A, Buczek P, Ernst A, Sandratskii L, Gross E. 2016.
\textit{Physical Review B} 94:014503

\bibitem{kotliar2006electronic}
Kotliar G, Savrasov SY, Haule K, Oudovenko VS, Parcollet O, Marianetti C. 2006.
\textit{Reviews of Modern Physics} 78:865

\bibitem{Fang+08}
Fang C, Yao H, Tsai WF, Hu J, Kivelson SA. 2008.
\textit{Phys. Rev. B} 77:224509

\bibitem{Fernandes+12}
Fernandes RM, Chubukov AV, J.~Knolle IE, Schmalian J. 2012.
\textit{Phys. Rev. B} 85:024534

\bibitem{Wetterich93}
Wetterich C. 1993.
\textit{Phys. Lett. B} 301:90

\bibitem{Metzner+12}
Metzner W, Salmhofer M, Honerkamp C, Meden V, Sch\"onhammer K. 2012.
\textit{Rev. Mod. Phys.} 84:299

\bibitem{Salmhofer99}
Salmhofer M. 1999.
Renormalization, an introduction.
Berlin, Heidelberg: Springer-Verlag

\bibitem{Benfatto+90}
Benfatto G, Gallavotti G. 1990.
\textit{Phys. Rev. B} 42:9967

\bibitem{Shankar91}
Shankar R. 1991.
\textit{Physica A (Amsterdam)} 177:530

\bibitem{Shankar94}
Shankar R. 1994.
\textit{Rev. Mod. Phys.} 66:129

\bibitem{Polchinski92}
Polchinski J. 1992.
\textit{Nucl. Phys., B} 422:617

\bibitem{Husemann+09}
Husemann C, Salmhofer M. 2009.
\textit{Phys. Rev. B} 79:195125

\bibitem{Raghu+08}
Raghu S, Qi XL, Honerkamp C, Zhang SC. 2008.
\textit{Phys. Rev. Lett.} 100:156401

\bibitem{Pena+17}
de~la Pe\~na DS, Lichtenstein J, Honerkamp C. 2016.
Competing electronic instabilities of extended hubbard models on the honeycomb
  lattice: A functional renormalization group calculation with high wavevector
  resolution.
ArXiv:1606.01124

\bibitem{Needs2010}
Needs R, Towler M, Drummond N, R{\'\i}os PL. 2009.
\textit{Journal of Physics: Condensed Matter} 22:023201

\bibitem{Whitehead16}
Whitehead T, Michael M, Conduit G. 2016.
\textit{Physical Review B} 94:035157

\bibitem{Pilati10}
Pilati S, Bertaina G, Giorgini S, Troyer M. 2010.
\textit{Physical review letters} 105:030405

\bibitem{Chang11}
Chang SY, Randeria M, Trivedi N. 2011.
\textit{Proceedings of the National Academy of Sciences} 108:51--54

\bibitem{Bugnion14}
Bugnion P, R{\'\i}os PL, Needs R, Conduit G. 2014.
\textit{Physical Review A} 90:033626

\bibitem{ma2012density}
Ma PN, Pilati S, Troyer M, Dai X. 2012.
\textit{Nature Physics} 8:601--605

\bibitem{si2001locally}
Si Q, Rabello S, Ingersent K, Smith JL. 2001.
\textit{Nature} 413:804--808

\bibitem{kotliar2000landau}
Kotliar G, Lange E, Rozenberg M. 2000.
\textit{Physical review letters} 84:5180

\bibitem{lee2006doping}
Lee PA, Nagaosa N, Wen XG. 2006.
\textit{Reviews of modern physics} 78:17

\bibitem{Senthil:2004qi}
Senthil T, Vishwanath A, Balents L, Sachdev S, Fisher M. 2004.
\textit{Science} 303:1490--1494

\bibitem{kuklov2008deconfined}
Kuklov A, Matsumoto M, Prokof'Ev N, Svistunov B, Troyer M. 2008.
\textit{Physical Review Letters} 101:050405

\bibitem{green2016feynman}
Green A, Hooley C, Keeling J, Simon S. 2016.
\textit{arXiv preprint arXiv:1607.01778}

\bibitem{read1989valence}
Read N, Sachdev S. 1989.
\textit{Physical Review Letters} 62:1694

\bibitem{maldacena1998adv}
Maldacena J. 1998.
\textit{Int. J. Theor. Phys} 38:1113

\bibitem{maldacena2005illusion}
Maldacena J. 2005.
\textit{Scientific American} 293:56--63

\bibitem{verlinde2011origin}
Verlinde E. 2011.
\textit{Journal of High Energy Physics} 2011:1--27

\bibitem{PhysRevB.15.1523}
Geldart DJW, Rasolt M. 1977.
\textit{Phys. Rev. B} 15:1523--1532

\bibitem{huang1957quantum}
Huang K, Yang CN. 1957.
\textit{Physical review} 105:767

\bibitem{lee1957many}
Lee T, Yang C. 1957.
\textit{Physical Review} 105:1119

\bibitem{abrikosov1958concerning}
Abrikosov A, Khalatnikov I. 1958.
\textit{Sov. Phys. JETP} 6:888

\end{thebibliography}

\end{document}